\documentstyle[12pt]{article}
\begin{document}
\thispagestyle{empty}

\newcommand{\be}{\begin{equation}}
\newcommand{\ee}{\end{equation}}
\newcommand{\sect}[1]{\setcounter{equation}{0}\section{#1}}
\renewcommand{\theequation}{\thesection.\arabic{equation}}
\newcommand{\vs}[1]{\rule[- #1 mm]{0mm}{#1 mm}}
\newcommand{\hs}[1]{\hspace{#1mm}}
\newcommand{\mb}[1]{\hs{5}\mbox{#1}\hs{5}}
\newcommand{\bea}{\begin{eqnarray}}
\newcommand{\ena}{\end{eqnarray}}

\newcommand{\wt}[1]{\widetilde{#1}}
\newcommand{\und}[1]{\underline{#1}}
\newcommand{\ov}[1]{\overline{#1}}
\newcommand{\sm}[2]{\frac{\mbox{\footnotesize #1}\vs{-2}}
		   {\vs{-2}\mbox{\footnotesize #2}}}
\newcommand{\prt}{\partial}
\newcommand{\eps}{\epsilon}

\newcommand{\R}{\mbox{\rule{0.2mm}{2.8mm}\hspace{-1.5mm} R}}
\newcommand{\Z}{Z\hspace{-2mm}Z}

\newcommand{\cd}{{\cal D}}
\newcommand{\cg}{{\cal G}}
\newcommand{\ck}{{\cal K}}
\newcommand{\cw}{{\cal W}}

\newcommand{\vj}{\vec{J}}
\newcommand{\vl}{\vec{\lambda}}
\newcommand{\vz}{\vec{\sigma}}
\newcommand{\vt}{\vec{\tau}}
\newcommand{\vw}{\vec{W}}
\newcommand{\poiss}{\stackrel{\otimes}{,}}

% REVUES POUR BIBLIO

\newcommand{\NP}[1]{Nucl.\ Phys.\ {\bf #1}}
\newcommand{\PL}[1]{Phys.\ Lett.\ {\bf #1}}
\newcommand{\NC}[1]{Nuovo Cimento {\bf #1}}
\newcommand{\CMP}[1]{Comm.\ Math.\ Phys.\ {\bf #1}}
\newcommand{\PR}[1]{Phys.\ Rev.\ {\bf #1}}
\newcommand{\PRL}[1]{Phys.\ Rev.\ Lett.\ {\bf #1}}
\newcommand{\MPL}[1]{Mod.\ Phys.\ Lett.\ {\bf #1}}
\newcommand{\BLMS}[1]{Bull.\ London Math.\ Soc.\ {\bf #1}}
\newcommand{\IJMP}[1]{Int.\ Jour.\ of\ Mod.\ Phys.\ {\bf #1}}
\newcommand{\JMP}[1]{Jour.\ of\ Math.\ Phys.\ {\bf #1}}
\newcommand{\LMP}[1]{Lett.\ in\ Math.\ Phys.\ {\bf #1}}

%\begin{document}
\renewcommand{\thefootnote}{\fnsymbol{footnote}}

\newpage
\setcounter{page}{0}
\pagestyle{empty}

\vs{30}

\begin{center}

{\LARGE {\bf Rational $\cw$ algebras from}}\\
{\LARGE {\bf composite operators}}\\[1cm]

\vs{10}

{\large F. Delduc$^1$, L. Frappat$^2$, P. Sorba$^{1,2}$, F.
Toppan$^2$}

{\em Laboratoire de Physique Th\'eorique }
{\small E}N{\large S}{\Large L}{\large A}P{\small P}
\footnote{URA 14-36 du CNRS, associ\'ee \`a l'Ecole Normale
Sup\'erieure de
Lyon, et au Laboratoire d'Annecy-le-Vieux de Physique des Particules
(IN2P3-CNRS),

\noindent
$^1$ Groupe de Lyon: ENS Lyon, 46 all\'ee d'Italie, F-69364 Lyon
Cedex 07,
France,

\noindent
$^2$ Groupe d'Annecy: LAPP, Chemin de Bellevue BP 110, F-74941
Annecy-le-Vieux Cedex, France.
}\\

\vs{3}

and

\vs{3}

{\large E. Ragoucy}\footnote{On leave of absence from Laboratoire de
Physique
Th\'eorique ENSLAPP.}

{\em NORDITA,

Blegdamsvej 17, DK-2100 Copenhagen \O, Denmark}

\end{center}
\vs{20}

\centerline{ {\bf Abstract}}

\indent

Factoring out the spin $1$ subalgebra of a $\cw$ algebra leads to a
new $\cw$ structure which can be seen either as a rational finitely
generated $\cw$ algebra or as a polynomial non-linear ${\cw}_\infty$
realization.

\vfill
\rightline{{\small E}N{\large S}{\Large L}{\large A}P{\small
P}-AL-429/93}
\rightline{NORDITA-93/47-P}
\rightline{June 1993}

\newpage
\pagestyle{plain}
\renewcommand{\thefootnote}{\arabic{footnote}}
\setcounter{footnote}{0}

\sect{Introduction}

\indent

Although a large set of $\cw$ algebras and superalgebras is today
available,
a complete classification of such new symmetries is not yet obtained
(for recent reviews see \cite{feh} and \cite{bou}). As far as the
classical case is concerned, the known $\cw$ algebras and
superalgebras can be constructed
as symmetries -- conserved currents --
of Toda and super-Toda theories \cite{{feh},{fra}}. Each such $\cw$
algebra is generated by a finite number of primary fields $W_{h_k}$
($k = 1,...,N)$ and is of polynomial type:
this means
that the Poisson brackets of two primary fields $W_{h_i}$ and
$W_{h_j}$ have the following structure
\be
\{W_{h_i}(z),W_{h_j}(w)\} = \sum_{\alpha}
P_{\alpha}(W_{h_k}(w),\prt^n
W_{h_k}(w)) \delta^{(\alpha)}(z-w)
\ee
the $P_{\alpha}$ being polynomials in the $W_{h_k}$ and their
derivatives.
Note that also the occurence of $\cw$ algebras in which the
polynomials are replaced
by rational functions in the $W_{h_k}$ and their derivatives, has
been discussed in \cite{feh2}.

Besides the finitely generated $\cw$ algebras, the $\cw_\infty$ ones
have also
been considered \cite{bak}. In this second class of $\cw$ algebras,
the most popular one -- and the oldest
one -- is the area-preserving diffeomorphism algebra on a two
dimensional manifold, also denoted $w_{\infty}$. Different
deformations of
$w_{\infty}$ have already
been obtained \cite{ara} and various realizations of such
$\cw_{\infty}$-algebras recently
considered (see for example \cite{pop} for a review). Such structures
show up rather
naturally in conformal field theoretical approaches of the quantum
Hall effect
\cite{cap}, of the black hole problem \cite{{ell},{yuw}} and also in
the matrix models approach \cite{bon}.

\indent

One of the purposes of this letter is to show that a large class of
rational
$\cw$ algebras can be obtained from finitely generated polynomial
$\cw$ algebras. Let us,
at this point, make precise or recall, some general definitions and
properties on finitely generated $\cw$ algebras. Indeed, there are
two basic ingredients
in the construction of a Toda theory \cite{lez}: a simple Lie algebra
-- or
superalgebra -- $\cg$ and an $Sl(2)$ -- or $OSp(1|2)$ -- subalgebra.
When $Sl(2)$ is chosen as the principal subalgebra of $\cg$, then one
gets
the $\cg$-abelian Toda model. The corresponding
$\cw$ algebra of conserved quantities is generated by primary
fields of conformal spin $h = 2,...,k$, these numbers being actually
the
degrees of the fundamental polynomial invariants of $\cg$. As an
example, for $\cg = Sl(3)$, one gets the algebra usually denoted
$\cw_3$ or $\cw(A_2)$
\footnote{We denote by $\cw(\cg,\ck)$ the $\cw$ algebra based on a
Toda model characterized by the Lie algebra $\cg$ and the $Sl(2)$
embedding principal in the subalgebra $\ck$ of $\cg$. In the case
of the abelian Toda model where $\ck = \cg$, the corresponding
$\cw$ algebra is simply denoted by $\cw(\cg)$.}
and generated by the fields $W_2, W_3$. By extension, we will call
such
algebras abelian $\cw$ algebras. When $Sl(2)$ is not principal in
$\cg$, the conserved quantities of the corresponding non abelian
Toda model will generate
what we will call a non abelian $\cw$ algebra. In $Sl(3)$,
there is only one $Sl(2)$ which is not principal in it, its Dynkin
index is 1.
The generators of the corresponding non abelian $\cw$ algebra
$\cw(A_2,A_1)$ are, in addition to $W_2$, two conformal
spin-$\sm{3}{2}$ fields
$W_{3/2}^{\pm}$ and a spin-one field $W_1$; this algebra is also
called the Bershadsky algebra \cite{ber}.

\indent

One can remark
\footnote{A study of $Sl(2)$ embeddings in semi-simple Lie algebras
$\cg$ leading to $\cw$ algebras without Kac-Moody component can be
found in ref. \cite{bow}.}
that most of the non abelian $\cw$ algebras contain spin-one fields.
It is reasonable to expect the spin-one part in a $\cw$ algebra to
play a particular role. One can easily
realize that these fields close linearly into a Kac-Moody algebra
$\cw_1$.
Note that the horizontal (or finite part) of $\cw_1$ is the commutant
of
the chosen $Sl(2)$ in $\cg$. Moreover, the $\cw$ generators decompose
into
irreducible representations under the adjoint action of this
Kac-Moody
algebra. It is also known that in a super $\cw$ algebra the spin 1/2
part, when it exists, can be factored out \cite{god}. More precisely,
a
meromorphic conformal
field theory can be decomposed into the tensor product of a spin 1/2
part and a conformal field theory without such spin 1/2 fields.
Similarly, one could wonder what will result from the factorization
in $\cw$ of its $\cw_1$ subalgebra.

As is shown below, rational $\cw$ structures appear when determining
the commutant $Com_{\cw}{\cw_1}$ of $\cw_1$ in $\cw$ algebras.
Each such a rational $\cw$ algebra provides an explicit realization
of a non-linear -- but polynomial -- $\cw_\infty$ algebra. This new
structure appears by assuming the primary fields,
which are expressed as rational functions of the basic fields in the
original rational $\cw$ algebra, to be independent algebra
generators.
Furthermore, the explicit construction of the fields in the commutant
 naturally involves a derivative which is covariant with respect to
 the
Kac-Moody
transformations generated by $\cw_1$.

After presenting the general properties of the commutant
$Com_\cw{\cw_1}$ when $\cw_1$ is restricted to be a $U(1)$
Kac-Moody, we work out explicitely
the simplest case, the one associated to $\cw(A_2,A_1)$ (the
Bershadsky algebra). Then we illustrate the situation for non-abelian
$\cw_1$'s with the specific examples taken from $A_3$ non-abelian
Toda models.

\sect{The U(1) case}

\subsection{General features}

\indent

We will consider in this section the case of a $\cw$ algebra
containing a spin-one
field. Let $J(z)$ be the spin-one primary field under the $T_0(z)$
Virasoro field in $\cw$. Their Poisson brackets are
\bea
&& \{T_0(z),T_0(w)\} = -2T_0(w)\delta'(z-w) + \prt T_0(w)\delta(z-w)
+ c\delta'''(z-w) \nonumber \\
&& \{T_0(z),J(w)\} = -J(w)\delta'(z-w) + \prt J(w)\delta(z-w)
\nonumber \\
&& \{J(z),J(w)\} = -\gamma\delta'(z-w)
\label{refeqn5}
\ena
The other primary fields $W^q_h(z)$ in $\cw$ undergo the $T_0(z)$
action as follows
\be
\{T_0(z),W^q_h(w)\} = -hW^q_h(w)\delta'(z-w) + \prt W^q_h(w)
\delta(z-w)
\label{refeqn6}
\ee
where $h$ denotes the conformal dimension, while $q$ specifies the
$U(1)$ charge carried by the primary field:
\be
\{J(z),W^q_h(w)\} = qW^q_h(w)\delta(z-w),
\ee
This last relation reflects the action of the Kac-Moody part on the
rest of the $\cw$ algebra in accordance with refs.
\cite{{fra},{bai}}.

A new stress-energy tensor can be defined by subtracting from
$T_0(z)$
a Sugawara term constructed with the $J(z)$ field, through the
relation
\be
T(z) = T_0(z) - \frac{1}{2\gamma} J(z)^2,
\ee
The first and third relations in (\ref{refeqn5}) remain valid when
shifting $T_0(z)$ into $T(z)$, while the second relation now gives
\be
\{T(z),J(w)\} = 0
\label{refeqn2}
\ee
Equation (\ref{refeqn6}) then gives
\be
\{T(z),W^q_h(w)\} = -hW^q_h(w)\delta'(z-w) + \cd W^q_h(w) \delta(z-w)
\label{refeqn7}
\ee
with
\be
\cd W^q_h(w) = (\prt - \frac{q}{\gamma} J(w))W^q_h(w)
\ee
The presence of a covariant derivative associated to the Kac-Moody
part of a
$\cw$ algebra has already been pointed out in ref. \cite{bai}. One
can directly verify that for any positive integer $n$
\be
\{J(z),\cd^n W^q_h(w)\} = q\cd^n W^q_h(w)\delta(z-w)
\ee
which insures that under the "inner automorphisms" generated by
$J(z)$
\bea
X(z) \rightarrow && X(z) + \int dz' \alpha(z') \{J(z'),X(z)\}
\nonumber\\
&& + \sm{1}{2} \int
dz' dz'' \alpha(z')\alpha(z'') \{J(z''),\{J(z'),X(z)\}\} + ...
\ena
the fields $\cd^n W^q_h(z)$ transform as
\be
\cd^n W^q_h(z) \rightarrow e^{q\alpha(z)} \cd^n W^h_q(z)
\ \ \ \mbox{with }
m=0,1,2,...
\label{refeqn8}
\ee
as must be the case for $\cd$ a covariant
derivative.
One should stress that the use of the covariant derivative is
particularly useful since it allows to write down in a compact form
the Poisson brackets as we will do in the following.

It is easily shown that the commutant $Com_{\cw}(J)$
of $J(z)$ -- that is in other words the polynomial $\cw$ subalgebra
with
elements having a zero Poisson bracket
with $J(z)$ -- is generated by $\partial^{n_0}T(z)$ and by the
monomials
\be
(\cd^{n_1} W^{q_1}_{h_1}) (\cd^{n_2} W^{q_2}_{h_2})...(\cd^{n_k}
W^{q_k}_{h_k})
\ee
where $n_0$ and the $n_i$'s are non negative integers, and the
charges $q_i$ satisfy the condition
\be
\sum_{i=1}^k q_i = 0
\ee

A remarkable feature of $Com_W(J)$ is that it contains an infinite
tower
of primary fields of integral dimension. The primary fields can be
specified
by their order $n$, which counts the number of covariant derivatives
in their leading term (subleading terms, having a lower number of
covariant derivatives, should
be added to get a primary field). A primary field of zeroth order
is just given by the product $W_h^0$:
\be
W_h^0(z) = \prod_{j=1}^k W_{h_j}^{q_j}
\ee
with
\be
\sum_{j=1}^k q_j = 0 \mbox{ and } \sum_{j=1}^k h_j = h
\label{refeqn9}
\ee
with $W_{h_j}^{q_j}$ primary in $\cw$ with respect to $T_0(z)$ (cf.
equation \ref{refeqn6}). Indeed, we can easily check that
\be
\{T(z),W_h^0(w)\} = -hW_h^0(w) \delta'(z-w) + \prt W_h^0(w)
\delta(z-w)
\ee
The primary fields of order $n$ obtained from $W_h^0$ have conformal
dimension $h+n$.
\\
A primary field at first order is
\be
W_{h+1}^0(z) = \sum_{j=1}^k \alpha_j \cd_j (\prod_{m=1}^k
W_{h_m}^{q_m})
\ee
where $\cd_j$ denotes the covariant derivative applied to the $j$th
term of the product, and the coefficients $\alpha_j$ are such that
\be
\sum_{j=1}^k \alpha_j h_j = 0
\ee

At second order, the primary fields are of the form
\be
W_{h+2}^0(z) = \sum_{i \le j = 1}^k \alpha_{ij} \cd_i \cd_j
(\prod_{m=1}^k W_{h_m}^{q_m}) + fTW_h^0
\ee
with the two sets of condition:
\be
-fc + \sum_{j=1}^k \alpha_{jj} h_j = 0
\ee
and
\be
\alpha_{ii} (2h_i + 1) + \sum_{j \ne i} \alpha_{ij} h_j = 0
\ \ \ \mbox{for } i=1,..,k
\ee
in addition to the null charge restriction eq. (\ref{refeqn9}$a$).
The coefficients $\alpha_{ij}$, as well as the $\alpha_{ijl}$
introduced below,
are assumed symmetric in the exchange of two indices.
\\
At third order, the primary fields are given by
\be
W_{h+3}^0(z) =
\sum_{i\leq j\leq l}^k \alpha_{ijl} \cd_i \cd_j \cd_l(\prod_{m=1}^k
W^{q_m}_{h_m}) + \sum_{j=1}^k g_j T \cd_j (\prod_{m=1}^k
W^{q_m}_{h_m}) + f(\prt T) W_h^0
\ee
The vanishing of the coefficients of the $\delta''$
term in the Poisson brackets implies the relations
\bea
\alpha_{iii}(3h_i+3) +\sum_{j\neq i} \alpha_{iij}h_j &=& 0\nonumber\\
\alpha_{iij}(2h_i+1)+\alpha_{jji}(2h_j+1) +\sum_{k\neq i,j}
\alpha_{ijk}h_k &=& 0 \ \ \ \mbox{for } i\ne j
\ena
and the vanishing of the coefficients of $\delta''',\delta''''$ leads
respectively to
\bea
cg_j -(3h_j+1)\alpha_{jjj} - \sum_{i\ne j}h_i\alpha_{iij} &=& 0
\nonumber\\
-fc + \sum_i^k\alpha_{iii}h_i &=& 0
\ena

A similar set of equations can be constructed at any order and it can
be immediately checked that it always admits solutions. Moreover,
at least for the invariants generated by bilinear products of two
fields
with opposite charges, the solution is unique.

\subsection{An example : the U(1) Commutant of the Bershadsky
algebra $\cw (A_2, A_1)$}

Let us treat for illustration the case of the Bershadsky algebra.
This algebra contains one spin-two field (the stress-energy tensor
$T(z)$), one spin-one
field (the current $J(z)$) and a couple of spin-$\sm{3}{2}$ fields
$W_{\pm}(z)$
of opposite charge under the $U(1)$ current. The commutation
relations are:
\bea
&& \{J(z),J(w)\} = - \sm{3}{2} c\delta'(z-w) \nonumber \\
&& \{T(z),T(w)\} = -2 T(w) \delta'(z-w) + \prt T(w) \delta(z-w) +
\sm{c}{2} \delta'''(z-w) \nonumber \\
&& \{T(z),J(w)\} = 0 \nonumber \\
&& \{T(z),W_{\pm}(w)\} = -\sm{3}{2} W_{\pm}(w) \delta'(z-w) + (\cd
W_{\pm})(w) \delta(z-w) \nonumber \\
&& \{J(z),W_{\pm}(w)\} = \pm \sm{3}{2} W_{\pm}(w) \delta(z-w)
\nonumber \\
&& \{W_+(z),W_-(w)\} = (T - c \cd^2)(w) \delta(z-w)
\nonumber \\
&& \{W_+(z),W_+(w)\} = \{W_-(z),W_-(w)\} = 0
\nonumber \\
&&
\ena

Notice that the Poisson brackets can be compactly written  using
the covariant derivative introduced in the previous section:
\bea
\cd W_\pm&=& (\partial \mp \sm{1}{c}J) W_\pm\nonumber
\ena

Clearly the commutant of the $U(1)$ current contains the
stress-energy
tensor $T$, its derivatives, and the bilinear products $W^{(p,q)}=
(\cd^p W_+)(\cd^qW_-)$,
with $p,q$ non-negative integers. As explained in the previous
paragraph, the
fields $W^{(p,q)}$ and $T$ are the building blocks from which one
constructs an infinite tower
of primary fields $W_{3+n}$ ($n\geq 0$) of integral dimension $ 3
+n$, one for each value of $n$; at the first orders we get, in a
specific normalization,
\bea
W_3 &=& W_+\cdot W_- \nonumber\\
W_4 &=& W_+\cdot\cd W_- - W_-\cdot \cd W_+ \nonumber \\
W_5 &=& \sm{3}{2} \cd^2W_+\cdot W_- - 4\cd W_+\cdot\cd W_- +
 \sm{3}{2} \cd^2W_-\cdot W_+ + \sm{9}{c} T\cdot W_3 \nonumber \\
W_6 &=& (\cd W_+)\cdot (\cd^2 W_-) - (\cd W_-)\cdot (\cd^2 W_+)
- \sm{1}{6} \prt^2 W_4 - \sm{13}{3c} T\cdot W_4
\label{primaryfields}
\ena
Analogous formulae hold for $n> 3$.

The primary fields $W_{3+n}$ with $n\geq 2$ are not independent: they
can be expressed as
rational functions of $T, W_3, W_4$ and their derivatives, as the
following reasoning shows. At first one can notice that $W^{(p,q)}$
for $q\geq 1$ are linear combinations of $V^{(p')}\equiv W^{(p',0)}$
and their derivatives; besides
that, due to the properties of the covariant derivatives, the
following quadratic relations hold:
\bea
V^{(p+1)} \cdot V^{(0)} &=& V^{(0)} \cdot \partial V^{(p)} - V^{(p)}
\cdot
\partial V^{(0)} + V^{(p)} \cdot V^{(1)}
\ena
They show that the fields $V^{(p')}$ ($p'\geq 2$) are obtained as
rational functions of
\bea
V^{(0)} &=& W_3 \nonumber\\
V^{(1)} &=& \sm{1}{2}(\partial W_3 -W_4)
\label{v0v1}
\ena

The ${\bf U(1)}$-Commutant algebra $Com_W(J)$ of the Bershadsky
algebra
has therefore the structure of a finitely generated rational
$\cw$-algebra of type $(2,3,4)$.
It is explicitely given by $T, W_3, W_4$ satisfying:
\bea
\{T(z),T(w)\} &=& -2 T(w) \delta'(z-w) + \prt T(w) \delta(z-w) +
\sm{1}{2}
\delta'''(z-w) \nonumber \\
\{T(z),W_{3+n}(w)\} &=& -(3+n) W_{3+n}(w) \delta'(z-w) + \prt
W_{3+n}(w) \delta(z-w) \quad \nonumber \\
\{W_3(z),W_3(w)\} &=& 2 W_4(w) \delta'(z-w) - \prt W_4(w)
\delta(z-w) \nonumber \\
\{W_3(z),W_4(w)\} &=& - 2 W_3(w) \delta'''(z-w) - 4 \prt W_3(w)
\delta''(z-w) \nonumber \\
&& + \sm{2}{7} [W_5 - 16 T\cdot W_3 + 9 \prt^2 W_3](w) \delta'(z-w)
\nonumber \\
&& -\sm{2}{7} \prt [W_5 - 16 T\cdot W_3 + 2 \prt^2 W_3](w)
\delta(z-w)
\nonumber \\
\{W_4(z),W_4(w)\} &=& 6 W_4(w) \delta'''(z-w) - 9 \prt W_4(w)
\delta''(z-w) \nonumber \\
&& + [-8W_6 + \sm{17}{3}\prt^2 W_4 + 6W_3\cdot W_3 -
\sm{116}{3}T\cdot W_4)](w)
\delta'(z-w) \nonumber \\
&& + \prt [4W_6 - \sm{4}{3}\prt^2 W_4 - 3W_3\cdot W_3 +
\sm{58}{3}T\cdot W_4)]
(w) \delta(z-w)\nonumber\\
&&
\label{commalgebra}
\ena
where $n$ is a non-negative integer; the explicit expression of the
primary
fields
$W_{5,6}$ in terms of $T, W_{3,4} $ can be computed from
(\ref{primaryfields},\ref{v0v1}). We get
\begin{eqnarray}
W_5 &=& {1\over 4W_3} (7\Psi +14W_3\partial W_4
-8W_3\partial^2 W_3 +36 T {W_3}^2 )\nonumber \\
W_6 &=& {-1\over 12{W_3}^2} (3W_4 \Psi + 3 W_3 \partial \Psi
+8{W_3}^2\partial^2 W_4 -6{W_3}^2\partial^3W_3 +52 T {W_3}^2
W_4)\nonumber\\
&&
\end{eqnarray}
Here $\Psi$ is a $8$-dimensional polynomial in $W_3, W_4$ and their
derivatives:
\begin{eqnarray}
\Psi &=& {W_4}^2-2W_3\partial W_4 -{(\partial W_3)}^2
+2W_3\partial^2W_3
\end{eqnarray}
The above formulae are given for the central charge $c=1$. In the
classical case
the algebra for any other value of the central charge can be obtained
by simply rescaling the fields and the Poisson brackets as well.

A few comments are in order: the algebra $Com_W(J)$ has a new
structure with respect to the standard $\cw$ algebras: its closure
is not on polynomials, but on rational functions of the fields and
their
derivatives. Such
algebra provides a specific realization of an underlying non-linear,
but of polynomial type, ${\cw }_\infty$ algebra, given by the Poisson
brackets of the primary
fields $W_{3+n}$ (and of $T$) among themselves, considered now as
independent algebra generators. Such a ${\cw }_\infty$ algebra is
non-linear as it can be immediately
seen from the relations (\ref{commalgebra}). Moreover it is a genuine
${\cw }_\infty$
algebra, i.e. it is {\it not} possible to find out a finite subset of
primary fields which, together with the stress-energy tensor, close
the
algebra {\it in a polynomial way}. The latter statement can be
immediately
verified by inspecting the Poisson brackets of the above introduced
monomials $W^{(p,q)}$ among themselves.
The situation here should be compared with that of standard finitely
generated ${\cw }$-algebras:
in that case \cite{pop2} to get rid of the non-linear character of
the
$\cw$ algebras one is
lead to introduce a linear ${\cw}_\infty$ algebra, promoting
non-linear terms to be new primary fields. Here, to get rid of the
non-polynomial character of the finitely generated rational
$\cw$ algebra, we promote the rationally expressed primary fields
to be new fields; the resulting ${\cw}_\infty$ algebra is in that
case non-linear.

Finally one should notice that the central charge is degenerated,
namely it
appears only in the Poisson brackets of the stress-energy tensor with
itself.

It is evident that, even if we have worked out explicitely the
construction
for the simplest example, the same structure holds in more general
cases.

\sect{The non abelian Kac-Moody case}

\indent

Let us now discuss the non abelian case. For this purpose, we will
treat two examples, namely the
algebra $\cw(A_3,A_1 \oplus A_1)$ associated to the WZNW reduction
$Sl(2)_2 \subset Sl(4)$ (the $Sl(2)$ subscript is the Dynkin index of
the
embedding) -- admitting a Kac-Moody subalgebra $\cw_1 = Sl(2)$ -- and
the algebra $\cw(A_3,A_1)$ associated to the WZNW reduction
$Sl(2)_1 \subset Sl(4)$ -- admitting a Kac-Moody subalgebra $\cw_1 =
Sl(2) \oplus
U(1)$. Finally, we will give some indications on the
general structure for the non abelian case.

\subsection{The example of $\cw(A_3,A_1 \oplus A_1)$}

\indent

Consider the $\cw$ algebra $\cw(A_3,A_1 \oplus A_1)$ generated by
three
spin-one fields $J^i(z)$ and four spin-two fields $W^i(z)$ and $T(z)$
with $i=1,2,3$. The Poisson brackets of this $\cw$ algebra are given
by
\bea
\{T(z),T(w)\} &=& -2T(w) \delta'(z-w) + \prt T(w) \delta(z-w)
+ c \delta'''(z-w) \nonumber \\
\{T(z),J^i(w)\} &=& 0 \nonumber \\
\{J^i(z),J^j(w)\} &=& c(\cd^{ij})_w  \delta(z-w)
 \nonumber \\
\{J^i(z),W^j(w)\} &=& - {\eps^{ij}}_k W^k(w) \delta(z-w) \nonumber \\
\{T(z),W^i(w)\} &=& -2W^j(w) \delta'(z-w) + (\cd W)^j(w)\delta(z-w)
\nonumber \\
\relax \{W^i(z),W^j(w)\} &=& -\sm{1}{2} ( ({\cd^3}_w)^{ji} -2 T
\cdot{\cd_w}^{ji} -\partial T\cdot \eta^{ji} )_w\delta (z-w)
\label{nonab}
\ena
where
$\eps^{ijk}$ is the completely antisymmetric tensor of
rank 3 such that $\eps^{123}=1$. The tensor $\eta_{ij}$
($\equiv \eta^{ij}$)
is the diagonal matrix $Diag(-1,1,1)$ and is used for lowering
(raising) the indices.\\
The covariant derivative is here
\bea
\cd^{ij} &=& \eta^{ij}\partial +\sm{1}{c} {\eps^{ij}}_k J^k
\ena

The transformations of the fields $J^i$ and $W^i$
with respect to the infinitesimal parameter $\lambda^i(z)$ are given
by
\bea
\delta W^i &=& {\eps^{i}}_{jk} \lambda^j W^k \nonumber\\
\delta J^i &=& c\partial \lambda^i +{\eps^i}_{jk}\lambda^j
J^k\nonumber
\ena
The application of the covariant derivative on $W^i$ leads to a
covariant
transformation property; in fact one gets ($n$ being an integer):
\bea
\{J^i(z),(\cd^nW)^j(w)\} &=& - {\eps^{ij}}_k (\cd^n W)^k(w)
\delta(z-w) \nonumber \\
\ena
and, in terms of infinitesimal transformations generated by
$\lambda^i$:
\bea
\delta (\cd^nW)^i &=& {\eps^{i}}_{jk} \lambda^j (\cd^n W)^k \nonumber
\ena

Notice that since the Virasoro generator $T$ commutes with the
Kac-Moody
fields
$J^i$, the covariant derivative $\cd T$ in equation (\ref{nonab}) has
to
be understood as $\prt T$.

The equations in (\ref{nonab}) are similar as those in
(\ref{refeqn2}) and (\ref{refeqn7}): they look formally just like
 the abelian case once the fields
are accomodated  into multiplets.  In particular the analysis
of the conformal dimensions of the derivatives fields $\cd^p\vw$ is
completely
analogous to that done in sect. 2 for the abelian case. Moreover,
(\ref{nonab}$b,d$) show that the fields
$W^i(z)$ are primary fields with conformal dimension 2 with respect
to the
Virasoro generator $T_0(z) = T(z) + \sm{1}{2}\vj^2(z)$.

Now, the covariant form of the $\cw$ algebra with respect to its
Kac-Moody
subalgebra
allows us to determine easily the non-linear $\cw_{\infty}$ commutant
of the Kac-Moody
subalgebra
$Sl(2)$ in the algebra $\cw(A_3,A_1 \oplus A_1)$. Indeed, one can
immediately check
that bilinear invariants can be obtained from
the scalar products $ ^t(\cd^p\vw)\cdot (\cd^q\vw)(w)$ (where the
upperscript $t$ denotes transposition); indeed we have
\bea
\{\vj(z) , \ ^t(\cd^p\vw)\cdot (\cd^q\vw)(w)\} = 0
\ena
Therefore, the elements in the commutant are generated as before by
$\partial^r
T$ and by
$ ^t(\cd^p\vw)\cdot (\cd^q\vw)(z)$, with $p,q,r$ non-negative
integers.

The primary fields in the commutant are given by the same formulas as
in the abelian case, with just the replacements
$W_+\rightarrow \vw$ and $W_-\rightarrow ^t\vw$. Notice that
$^t\vw\cdot\vw$
has the same conformal dimension $(h=3)$ as the commutant in the
Bershadsky
algebra.
\subsection{The example of $\cw(A_3,A_1)$}

\indent

Now, we consider the example of the $\cw$ algebra $\cw(A_3,A_1)$
generated
by four spin-one fields $J^i(z)$ ($i=1,2,3$) and $J^0(z)$, four
spin-$\sm{3}{2}$ fields $^t \vw_{\pm}(z) =
(W^u_{\pm}(z),W^d_{\pm}(z))$
(indices $u$ and $d$ stands for $up$ and $down$) and one
spin-two field $T(z)$. The Poisson brackets of this $\cw$ algebra
are given by
\bea
&& \{T(z),T(w)\} = -2T(w) \delta'(z-w) + \prt T(w) \delta(z-w)
+ c \delta'''(z-w) \nonumber \\
&& \{T(z),\vj(w)\} = \{T(z),J^0(w)\} = 0 \nonumber \\
&& \{J^i(z) , J^j(w)\} = {\eps^{ik}}_k J^k(w) \delta(z-w)
- \sm{1}{2} \delta'(z-w) \nonumber \\
&& \{J^0(z),J^0(w)\} = -\delta'(z-w) \nonumber \\
&& \{\vj(z),J^0(w)\} = 0 \nonumber \\
&& \{T(z),\vw_{\pm}(w)\} = -\sm{3}{2} \vw_{\pm}(w) \delta'(z-w)
+ (\cd \vw_{\pm})(w) \delta(z-w) \nonumber \\
&& \{J^i(z),\vw_{\pm}(w)\} = \sigma^i \vw_{\pm}(w)
\delta(z-w) \nonumber \\
&& \{J^0(z),\vw_{\pm}(w)\} = \pm \vw_{\pm}(w) \delta(z-w) \nonumber
\\
&& \{^t \vw_+(z) \poiss \vw_+(w)\} = \{^t \vw_-(z) \poiss \vw_-(w)\}
=  0
\nonumber \\
&& \{^t \vw_+(z) \poiss \vw_-(w)\} = - \sm{1}{2} [\cd^2 - T](w)
\delta(z-w)
\ena
where the $\sigma_i$ are the Pauli matrices satisfying the algebra
\bea
 [\sigma^i,\sigma^j]&=& 2{\eps^{ij}}_k\sigma^k\nonumber
\ena
and $\cd =\partial -\vj \cdot \vz -qJ^0 $ is the covariant
derivative.
The commutant is in this case generated by
$\partial^r T$ and by
$ ^t(\cd^p\vw_+)\cdot (\cd^q\vw_-)(z)$, with $p,q,r$ non-negative
integers.
The structure is again similar to the one given in the previous
examples, and in particular even in this case
the first composite invariant operator $^t(\vw_+)\cdot (\vw_-)(z)$
has conformal dimension equal to $3$.

\indent

{\Large {\bf Conclusions}}

\indent

We have shown that, as soon as a classical $\cw$ algebra admits a
Kac-Moody part, the set of fields commuting with the spin $1$ part
generates
a rational $\cw$ subalgebra. This one can also be seen as a
realization of a non-linear, but polynomial, ${\cw}_\infty$-algebra.

The exhibited structures
deserve more detailed studies. In particular the quantum version has
to be considered; in this case it is reasonable to expect the central
charge to play a crucial role.
Another interesting question concerns the possible exploitation of
the covariant derivatives extensively used in this approach for the
construction of $2$-dimensional integrable models.

\end{document}